\documentclass{elsart}

\usepackage{graphicx}
\graphicspath{{./images/}}
\usepackage{cite}
\usepackage{amsmath}
\usepackage{amssymb}
\usepackage[latin1]{inputenc}

\begin{document}

\begin{frontmatter}

 \title{Magnetism of 3d Transition Metals and Magnetic Anisotropy of one-dimension Ni Chains on Au(110)-(1$\times2$):
A Noncollinear $ab-initio$ Density-Functional-Theory Study}

\author[a]{Wei Fan},
\author[b,a]{Xin-Gao Gong}
\address[a]{ Key Laboratory of Materials Physics,
 Institute of Solid State Physics, Chinese Academy of Sciences, 230031-Hefei, P. R. China}
\address[b]{Department of Physics, Fudan University, 200433-Shanghai, P. R. China}

\date{\today}

 \begin{abstract} Based on Density Functional Theory (DFT)
 with non-collinear-magnetism formulations, we have calculated
 the magnetism of single atom of 3d transition metals and the magnetic anisotropy of supported Ni chains
 on Au(110)-(1$\times2$) surface. Our results show that surface relaxations enhance the orbital moments
 of left-end elements (Ti,V) and quench the orbital moments of right-end elements (Fe,Co,Ni) on the
 Au(110)-(1$\times$2) surface. This is because Ti and V atoms raise their positions above the trough of
 the reconstructed Au(110) surface and Fe, Co, Ni atoms trap more deeply in the trough after the surface
 relaxations.  From the study of magnetism of Ni one-dimensional atomic chains, their magnetic anisotropy
 are closely relate to their orbital quenching. The easy magnetized axis changes from the direction
 parallel to the chains to the direction perpendicular to the Ni chains when they absorb on the surface.
 These one-dimensional Ni chains in the trough of the reconstructed Au(110) surface have ferromagnetic
 order and are favored with chain-length less than 6.
\end{abstract}


\end{frontmatter}

\section{\label{Intro} Introduction}

 The high-density magnetic recording and the memory devices require the strong magnetic anisotropy.
 The low-dimensional materials due to the
 reduced dimensionality generally have a favorable direction and the physical properties along this
 direction are generally different from the other directions. Thus we have the most possibilities to
 find the strong magnetic anisotropy materials from these low-dimensional materials. On the other hand,
 the reduced dimensionality reduces the atomic coordination number and enhances the spin and orbital
 magnetic moments of the materials. Since the pioneering experiments for the magnetism of Fe strips on
 the W(110) and Cu(111) surfaces~\cite{Elmers1,Shen1,Hauschild1,Pratzer1}, the researches along this
 direction have been extended to others quasi-one-dimensional systems such as the monatomic magnetic
 chains on vicinal surface~\cite{Gambardella1,Gambardella2,Gambardella3} and Co magnetic dot-chain on
 Ru(0001) surface~\cite{Li1}. The transition from in-plane to out-of-plane magnetic anisotropy had been
 found in Fe nano-structure when it is approach to one-dimensional limit~\cite{Boeglin1}. Besides the
 celebrating properties such as the strong magnetic anisotropy, the enhanced spin and orbital moments,
 some new phenomena have been found, such as the temperature and time dependent
 magnetization~\cite{Elmers1,Shen1} which make the realistic applications of these novel experimental
 devices face new serious problem such as the instability of magnetic structure.

 It is valuable to investigate theoretically the magnetic stability of surface-supported novel
 structures~\cite{Dreysse1}. Many theoretical and computational methods have been used to studied the
 magnetism of one-dimensional structure on surface such as the
 KKR~\cite{Bellini1,Wildberger1,Lazarovits1,Eisenbach1}, and TB-LMTO~\cite{Komelj1} and PAW
 methods~\cite{Spisak1,Spisak2} based on Density Functional Theory, and the numerically tight-binding
 self-consistent calculation~\cite{Dovila1} and analytically tight-binding calculation~\cite{Druzinic1}.
 Some common arguments have been found such as the enhancement of surface magnetism, the magnetic
 anisotropy are close related to the orbital moments and the spin-orbit coupling interaction (SOC).

 The orbital polarization~\cite{Brooks1,Eriksson1,Solovyev1} can remedy the small orbital moments in the
 calculations of density functional theory to approach the experimental values, and especially
 reproduce the correct phase transition in Ce companying with the volume collapses~\cite{Eriksson1}.
 The orbital polarization plays the opposite role to crystal field which quenches the orbital
 moment to small value in crystal due to the splitting of ground state energy levels.
 However there are still many contradictions among the already known theoretical results. The
 density functional
 calculations with OP (orbital polarization) + SOC (spin-orbit coupling) have given the correct orbital
 moments but too large MAE (Magnetic Anisotropy Energy) for tetragonal and cubic Ni. If the OP term
 switches off the theoretical value of the magnetic anisotropy energy is close its experimental
 values~\cite{Hjortstarn1}. The magnetic anisotropy energy of CoPt calculated using
 the so-called c-RPA~\cite{Solovyev2} are weakly dependent on the new introduced OP interaction.
 The magnetic anisotropy energies are surprisingly consistent with experiments even only
 under LSDA approximation without the OP effects.
 The orbital moments of transition metals are well
 consistent with experiments. This method has been combined with
 PAW method to calculate quasi-particle spectrum of
 materials~\cite{Fuchs1}.

 Based on Brook's theory, the magnetic anisotropy is induced by the spin-orbital
 interaction~\cite{Brooks2}.  Although the orbital moments are underestimated, it's still valuable
 to study the magnetic anisotropy properties in the absence of (OP) orbital polarization term.
 Theoretical works based on Density Functional Theory have also discovered others interesting
 properties of surface supported magnetic chains. Such as in Ref.~\cite{Spisak2}, the authors have
 studied the magnetism of the ultrathin wires of fifth and sixth row elements supported on Cu(117)
 and Ag(117) vicinal surfaces. They found that only metals with a half-filled $d$ band are found to
 have magnetic order on Cu(117) surface, additionally the ferromagnetic order is energetically stable.
 On the contrary, on the Ag(117) surface, the anti-ferromagnetic order is stable. The magnetism of
 mono-atomic chain is strongly dependent on their local environment absorbed on the surfaces.

 Besides the high-index vicinal surfaces such as Ag(117), the reconstructed Au(110)-(1$\times$2)
 surface decorated with one-dimensional troughs along the closed-packed A-[1$\bar{1}$0] direction
 can be used as a template to grow one-dimensional nano-structure.
 Experimentally, the distribution of lengths of Ni chains shows
 that most Ni chains are short than 6~\cite{Hitzke1}. Our
 molecular simulation additionally shows that the lattice misfit between Ni
 chains and the Au substrate can induce the instability of Ni chains and
 makes them break into short segments~\cite{Fan1}.

 In this work we study the magnetism of mono-atomic Ni chains supported on Au(110)-(1$\times$2)
 surface using the same theoretical method as in Ref.~\cite{Spisak2}.
 Our studies focus on the magnetic anisotropy
 of surface-supported magnetic atomic chains. We perform the non-collinear magnetic calculations
 including the spin-orbit coupling interaction but without orbital polarization interaction.
 We have calculated the magnetic anisotropy energies of the supported Ni chains and found an
 off-plane easy magnetization axis perpendicular to the chains and the surface. On the contrary,
 the easy magnetization axes of the free-standing Ni chains are parallel to these chains.
 The change of the easy magnetization axes of the surface-supported chains are also found in the
 others systems~\cite{Dovila1}. Our results also indicate the closed relationship between the
 magnetic anisotropy and the orbital magnetism.

 The paper is arranged as follow. In the next section, we introduce the theoretical methods used in
 this paper. The third section will present the results for single 3d transition metal atoms absorbed
 on Au(110)-(1$\times$2) and Au(111) surface. The magnetism of a single magnetic atom
 on a surface is important to understand the magnetism of materials with more complex structures such
 as the one-dimensional nano-structure. The reduced dimensionality and coordination number for surface
 still enhance the magnetism of single absorbed atoms~\cite{Gross1,Gross2,Hanf1,Riegel1,Ortega1}.
 The results of short one-dimensional mono-atomic Ni chains are included in the forth section. Our studies
 focus on the magnetic anisotropy of Ni mono-atomic chains. Finally, we conclude our results.

 \section{\label{methods} Theoretical Methods: Non-collinear Density Functional Theory and PAW methods}

 We have calculated magnetism of Ni$_{n}$(n=1-5) chains supported on the Au(110)-(1$\times$2) surface
 based on Density Functional Theory~\cite{Hohenberg1,Kohn1} and the Methods of Projection of
 Augmentation Wave (PAW)~\cite{Blochl1} with the plane-wave base set and GGA Perdew-Burke-Ernzerhof's
 exchange-correlation potential~\cite{Perdew1}. The PAW methods used in VASP
 program~\cite{Kresse1,Kresse3,Hobbs1} is as accurate as frozen-core all-electron methods and
 hopefully improved to include all-electron relaxation. Our calculations include the non-collinear effects
 and the spin-orbit coupling which is proved important to heavy metals such as gold. In non-collinear
 formula, a magnetic moment as a vector can point to any direction in space. The orbital polarization (OP)
 interaction isn't included in our calculations.

 The basic theory used in this paper is Hobbs's fully unconstrained non-collinear DFT formulas based on
 plane wave basis sets and all-electron PAW methods~\cite{Hobbs1,Hobbs2}. Instead of distributing local
 quantization axes for every atoms, which is used in programs based on atom sphere approximation (ASA) or
 others analogous methods~\cite{Sandratskii1}, Hobbs's theory adopts only one global quantization axis
 and the vectors of magnetic electronic density are varying smoothly and continuously, which is more suited
 for the calculations of itinerant magnetism in transition metal materials.

 In non-collinear density functional theory~\cite{Hobbs1,Hobbs2,Sandratskii1}, the total energy is the
 functional of a
 density matrix $n_{\alpha\beta}$ where $\alpha$ and $\beta$ are the spin index along a defined
 quantization axis. $\alpha=\uparrow$ or $\downarrow$ represent the spin angular momentum point to the
 positive direction or negative direction of the quantization axis. The sum of the diagonal elements is
 the total charge density,
 that is $n_{_{Tr}}=\sum_{\alpha}n_{\alpha\alpha}$. In a non-collinear magnetic system, the off-diagonal
 elements are non-zero. The 2$\times$2 matrix can be expanded using the complex matrix basis
 ($I$, $\sigma_{x}$, $\sigma_{y}$, $\sigma_{z}$) and expressed as
   \begin{eqnarray}
   n_{\alpha\beta}=[n_{_{Tr}}\delta_{\alpha\beta}+\vec m
   \cdot\vec\sigma_{\alpha\beta}]/2 ,
   \end{eqnarray}
   \noindent where $\vec m=\sum_{\alpha\beta}n_{\beta\alpha}\cdot\vec\sigma_{\alpha\beta}$ is
   the magnetic electronic density and $\vec\sigma$ is the Pauli
   matrix. In PAW methods, the pseudo-density matrix can be
   expressed using the pseudo-wave functions as
    \begin{eqnarray}
    \tilde n_{\alpha\beta}(\vec{r})=\sum_{n}f_{n}\langle\tilde\Psi_{n}^{\beta}|\vec{r}\rangle
     \langle\vec{r}|\tilde\Psi_{n}^{\alpha}\rangle ,
    \end{eqnarray}
    where $f_{n}$ is the occupation number. The total electronic density matrix
    $n_{\alpha\beta}=\tilde n_{\alpha\beta}+n^{1}_{\alpha\beta}-\tilde n^{1}_{\alpha\beta}$,
    where $n^{1}_{\alpha\beta}$ and  $\tilde n^{1}_{\alpha\beta}$
    are the on-site electronic density matrixes. The total energy as the sum of three parts
    $E=\tilde{E}+E^{1}-\tilde{E}^{1}$, where $\tilde E$ the smooth functional of the
    pseudo-density matrix $\tilde n_{\alpha\beta}(\vec{r})$ is evaluated on a regular grid,
    $E^{1}$ and $\tilde{E}^{1}$ the functionals of on-site electronic density matrixes
    $n^{1}_{\alpha\beta}$ and  $\tilde n^{1}_{\alpha\beta}$ are calculated on radical support
    grid around every atom.

    The Kohn-Sham Hamilton and  Kohn-Sham equation can be obtained from the variation of
    total energy to the soft pseudo-density matrix and written as
     \begin{eqnarray}
      H^{\alpha\beta}\left[n\right]&=&-\frac{1}{2}\Delta \delta^{\alpha\beta}+\tilde{v}^{\alpha\beta}_{eff}
      \\ \nonumber
      &+& \sum_{(ij)}|\tilde{p}_{i}\rangle(\hat{D}^{\alpha\beta}_{ij}+^{1}D^{\alpha\beta}_{ij}
      -^{1}\tilde{D}^{\alpha\beta}_{ij})\langle\tilde{p}_{j}| ,
      \\ \nonumber
      \sum_{\beta}H^{\alpha\beta}|\tilde{\Psi}^{\beta}_{n}\rangle&=&
      \varepsilon_{n}S^{\alpha\alpha}|\tilde{\Psi}_{n}^{\alpha}\rangle .
 \end{eqnarray}
 \noindent where $S^{\alpha\alpha}$ is the overlapping operator,
 $\tilde{D}^{\alpha\beta}_{ij}$,~$^{1}\hat{D}^{\alpha\beta}_{ij}$ and $^{1}D^{\alpha\beta}_{ij}$ are
 the nonlocal interaction obtained by the PAW transformation, which are equivalent to the nonlocal
 pseudo-potential in the ultrasoft-potential methods.
 In non-collinear DFT formulas, the one-electron effective potential
 $\tilde{v}^{\alpha\beta}_{eff}=v_{H}(r)\delta_{\alpha\beta}+v_{xc}(r)\delta_{\alpha\beta}
 +\vec{b}(r)\cdot\vec{\sigma}^{\alpha\beta}$,
 which is the summation of electrostatic potential $v_{H}(r)$,
 the nonmagnetic part of GGA exchange-correlation correction $v_{xc}(r)$ and the magnetic
 exchange-correlation potential $\vec{b}(r)\cdot\vec{\sigma}^{\alpha\beta}$,
 where $\vec{b}=\delta E_{xc}/\delta \vec{m}(r)$.
 The magnetic exchange-correlation potential will potentially
 enhance magnetic moments in non-collinear calculations.

 The spin-orbit coupling enters into PAW formulas by all-electron
 part of PAW Hamiltonian and is expressed as
 \begin{eqnarray}
 H_{SOI}^{\alpha\beta}=\frac{\hbar^{2}}{(2m_{e}c)^{2}}\sum_{ij}
 \langle\phi_{i}|\frac{1}{r}\frac{dV_{sphere}}{dr}|\phi_{j}\rangle
 |\tilde{p}_{i}\rangle\vec{\sigma}_{\alpha\beta}\cdot\vec{L}_{ij}
 \langle\tilde{p}_{j}|.
 \end{eqnarray}
 where $\tilde{p}_{i}$ is the projector function and $\phi_{i}$ all-electron wave function.
 The one-electron effective potential is modified
 $\tilde{v}^{\alpha\beta}_{eff}\rightarrow\tilde{v}^{\alpha\beta}_{eff}+\tilde{H}_{SOI}^{\alpha\beta}$,
 and the same time, one part of nonlocal PAW potentials is modified according to
 $\hat{D}^{\alpha\beta}_{ij}\rightarrow\hat{D}^{\alpha\beta}_{ij}=\sum_{L}
 \int(\tilde{v}^{\alpha\beta}_{eff}(r)+\tilde{H}_{SOI}^{\alpha\beta}(r))\hat{Q}_{ij}^{L}(r)dr$.
 The roles of spin-orbit interaction in $^{1}\hat{D}^{\alpha\beta}_{ij}$ and $^{1}D^{\alpha\beta}_{ij}$
 are similar to $\hat{D}^{\alpha\beta}_{ij}$, however they are calculated only within atom
 sphere. In the calculations of real material, they adopt the values in the calculations of isolate atom and
 don't update in Kohn-Sham self-consistent loop.
 More information on non-collinear density functional
 theory within PAW methods can be found in the reference~\cite{Hobbs1}.

 The atoms can have their initial magnetic moments by constructing the magnetic electronic density and
 the magnetic moments around each of atoms are equal to their desired values.
 If the initial total magnetic moments point $\vec{n}$ direction, generally the direction of total moment
 changes in Kohn-Sham self-consistent loop. We can use the constrained DFT calculations to fix
 the direction of magnetic moments.
 Our self-consistent calculations show that the directions of magnetic moments change very small in
 ferromagnetic materials although using the unconstrained DFT calculations (see the table.(\ref{table2})
 in section~\ref{Ni_chain_15} ). In real ferromagnetic materials, the directions of magnetic moments
 are hard to change, once the magnetic domains have formed. This is because the change of magnetization
 direction must include the changes of magnetic moments of all magnetic atoms in the materials.
 An external magnetic field can force to change the direction of magnetic moments. Of course there
 exists an easy magnetization axis which is stable direction of magnetic moments.
 We need the 'strongest' magnetic field to change direction of magnetic
 moments away from the direction of the easy magnetization axis. However for the materials with
 more complex magnetic structure such as the canted and spiral magnetic structures, the relaxations
 of directions of magnetic moments are intrinsic, the use of unconstrained DFT is essentially
 important~\cite{Hobbs1,Hobbs2}.

 \section{\label{mag_3d} The magnetism of 3d transition metal atoms absorbed on
 Au(110)-(1$\times$2) and Au(111) surface}

 The missing row Au(110) surface reconstructs from Au(110) surface by having missed closed-packed Au chains
 every others, and forms one-dimensional trough along the close-packed A-[1$\bar{1}$0] direction.
 We construct a crystal slab with 4 atomic layers including 28 gold atoms. At the first step we optimized
 the lattice constant of gold crystal which is about 4.2$\AA$. In the procession of optimization,
 the RMM-DIIS algorithm~\cite{Kresse1} is applied to the total-energy minimum with plane-wave energy
 cutoff 229.9eV and the 6x6x6 Monkhorst-Pack K points mesh. Now the size of super-cell is
  11.8$\AA$$\times$8.4$\AA$$\times$5.94$\AA$.

\begin{figure}
 \begin{center}
 \includegraphics{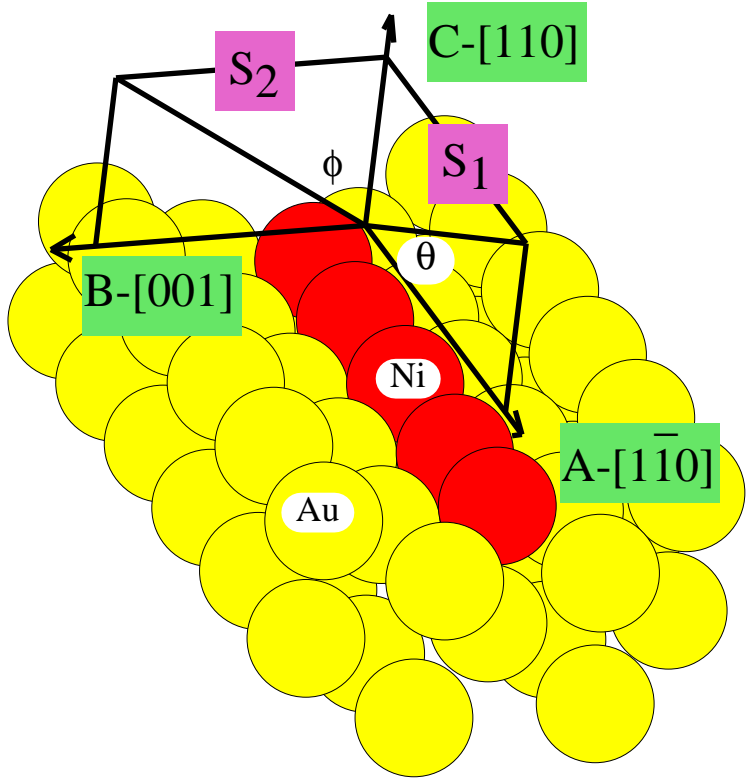}
 \caption{\label{fig1} The schematic diagram of a Ni chain supported on the Au(110)-(1$\times$2)
 surface. The trough of the reconstructed surface is along the A-[1$\bar{1}$0] direction.
 The red spheres are the absorbed
 Ni atoms in the trough. The yellow spheres are the Au atoms of the surface. The plane including both
 C-[110] axis and A-[1$\bar{1}$0] axis is named as S$_{1}$, the plane including both C-[110] axis and
 B-[001] axis as S$_{2}$. The rotating angle from any vector in S$_{1}$ to A-[1$\bar{1}$0] axis
 labels as $\theta$ and the rotating angle from any vector in S$_{2}$ to C-[110] axis labels as
 $\phi$. $\theta$ and $\phi$ are the magnetization angles in the two
 planes.}
 \end{center}
 \end{figure}

 We add a vacuum layer by extending the super-cell along C-[110]
 direction. The size along C-[110] direction of super-cell including vacuum layer is equal to 16$\AA$.
 The lattice constant of surface is generally smaller than that of the crystal. Thus we optimized the slab
 with vacuum layer together. Based on the optimized lattice constant, the size of the super-cell is
 11.523$\AA$$\times$8.148$\AA$$\times$15.52$\AA$. The vacuum layer is about twice thicker than thickness of
 the slab. For the optimized Au(111) surface, the size of super-cell is about
 11.5$\AA$$\times$9.97$\AA$$\times$20$\AA$ including 64 gold atoms. In above optimizations, we only change
 the lattice constant. The initial structures are constructed by placing the 3d transition metal atoms on
 the hollow sites of the Au(111) and Au(110)-(1$\times$2) surface in the trough (Fig.~\ref{fig1}). The initial
 structures are optimized for all atoms except for the bottom atoms by having used the conjugate-gradient
 Methods. The  K-points meshes of 2$\times$2$\times$1 or $\Gamma$ point are used to sampling the first
 Brillouin-zone in the corresponding calculations of the electronic structure for Au(110)-(1$\times$2)
 and Au(111) surfaces respectively. In this work we use smaller k-points mesh to more efficiently
 optimize surface structures.

 We use both collinear and non-collinear density functional calculations but mainly present the non-collinear
 results. In non-collinear DFT calculation, we use the optimized structures from collinear DFT calculation.
 Based on the relaxed structures, we calculate more exactly the electronic structures using the RMM-DIIS
 algorithm~\cite{Kresse1} with the (6$\times$6$\times$1) Monkhorst-Pack grids sampling the first Brillouin zone.
 If the changes of the total energies are smaller than 0.0001eV between two electronic self-consistent (SC)
 steps the SC-loops break. We set the Methfessel-Paxton smearing width equal to 0.20eV to accelerate the
 speed of convergence. The energy-cutoffs of plane-wave for different atoms are summarized in
 table~(\ref{table1}). The magnetic moments
 for a single atom are calculated by considering the electrons in a Wigner-Seitz sphere centered at the
 position of the atom. The Wigner Seitz Radii for 3d transition metal atoms are also summarized in
 table~(\ref{table1}). We choose the C-[110] direction as the quantization axis. We emphasize the
 influence of surface deformation on the orbital moments of the absorbed atoms. The orbital polarization
 interaction omits in this work but includes the spin-orbit coupling interaction. We obtain smaller
 orbital moments compared with pervious density functional calculation including the orbital polarization
 interaction~\cite{Nonas1}.

 \subsection{\label{mag_3d_relax} The influence of structural relaxations on the magnetism of
 a single absorbed 3d transition atom}

\begin{figure}
 \begin{center}
 \rotatebox{0}{\includegraphics{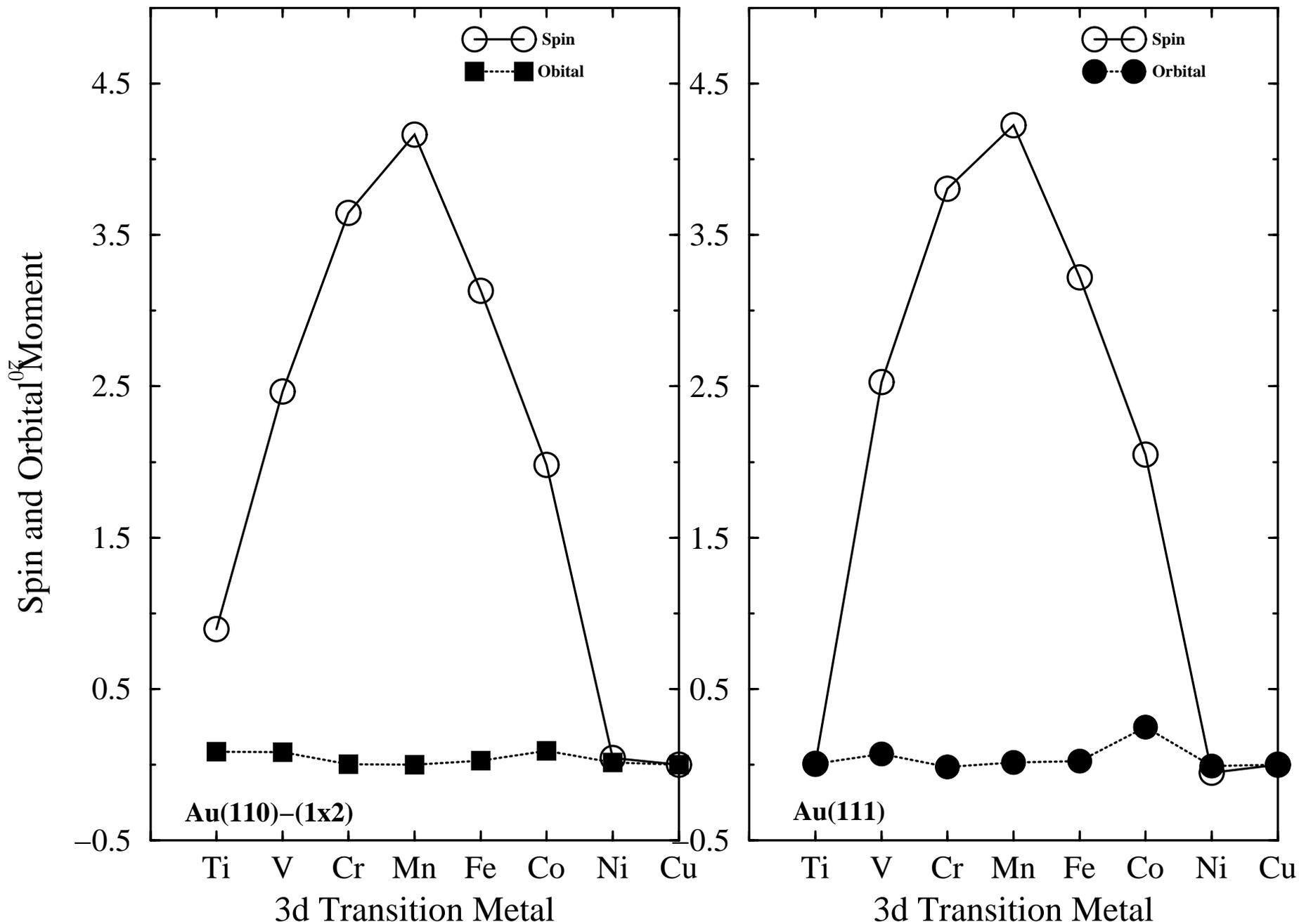}}
 \caption{\label{fig2} The spin (solid lines) and orbital (dot lines) moments of the absorbed 3d atoms. The left panel
  shows the results for Au(110)-(1$\times$2) surface, the right panel for the Au(111) surface.}
  \end{center}

 \end{figure}

 It's seem that collinear DFT calculation is enough to the study of magnetism of a single transition metal
 absorbed on the gold surface. However, if we want to know the direction of magnetic moments relative to
 the surface, we must use the non-collinear DFT formulas. These results for single absorbed atom will
 be compared with the results of one-dimensional atomic chains.
 At first step, we calculate the magnetic moments of the absorbed 3d atoms on the Au(110)-(1$\times$2) and
 Au(111) surfaces including the spin-orbit coupling interactions and non-collinear calculations.
 From Fig.~\ref{fig2}, we can see that the spin moments reach the maximum in the middle of the group.
 The large spin moment of the absorbed Mn atom is about 4.163$\mu_{_{B}}$ on the Au(110)-(1$\times$2)
 surface. We can see from the left panel of the figure that the spin moment of the absorbed Cr atom is
 the second largest when including the spin-orbit coupling interactions and non-collinear calculations.
 The change of magnetic moments per absorbed magnetic atom across the 3d row has been found in many
 other {\it ab initio} calculations such as 3d atoms on Au(001), Ag(001)~\cite{Cabria1} and
 Cu(001)~\cite{Stepanyuk1} surfaces. Our results are consistent very well with these results although
 the energy cutoffs of plane-waves are not too large in our calculations.

 Both the electronic correlations and the crystal field have significant influence on the orbital
 moments of the absorbed 3d atoms. If the electronic correlation is stronger than the crystal fields,
 the orbital moment is large, otherwise small. The absorbed Ti,V,Co atoms have visible orbital moments
 0.086$\mu_{_{B}}$, 0.082$\mu_{_{B}}$ and 0.092$\mu_{_{B}}$ on the Au(110)-(1$\times$2) surface, and
 0.006$\mu_{_{B}}$, 0.071$\mu_{_{B}}$ and 0.247$\mu_{_{B}}$ on the Au(111) surface respectively.
 Cabria, et.al obtained
 larger orbital moment (about 0.5$\mu_{_{B}}$) of Fe and Co atoms absorbed on Au(001) surface using the
 spin polarization relativistic KKR methods including the orbital polarization term~\cite{Cabria1,Nonas1}.
 The usual exchange-correlation potentials (LDA or GGA) underestimate the electronic correlations such as
 the Coulomb correlation and the orbital polarization, thus the obtained orbital moments in this work are
 generally small compared with experimental values~\cite{Brooks1,Solovyev1,Eriksson1,Nonas1} although our
 work includes the spin-orbit coupling interaction. The small values of orbital moments of the absorbed
 Cr atoms in our calculations are not related to the crystal field but to its electronic structure. In the
 individual Cr atom, five 3d-electrons half-fill the 3d states, the total orbital moment is very small.

\begin{figure}
 \begin{center}
 \includegraphics{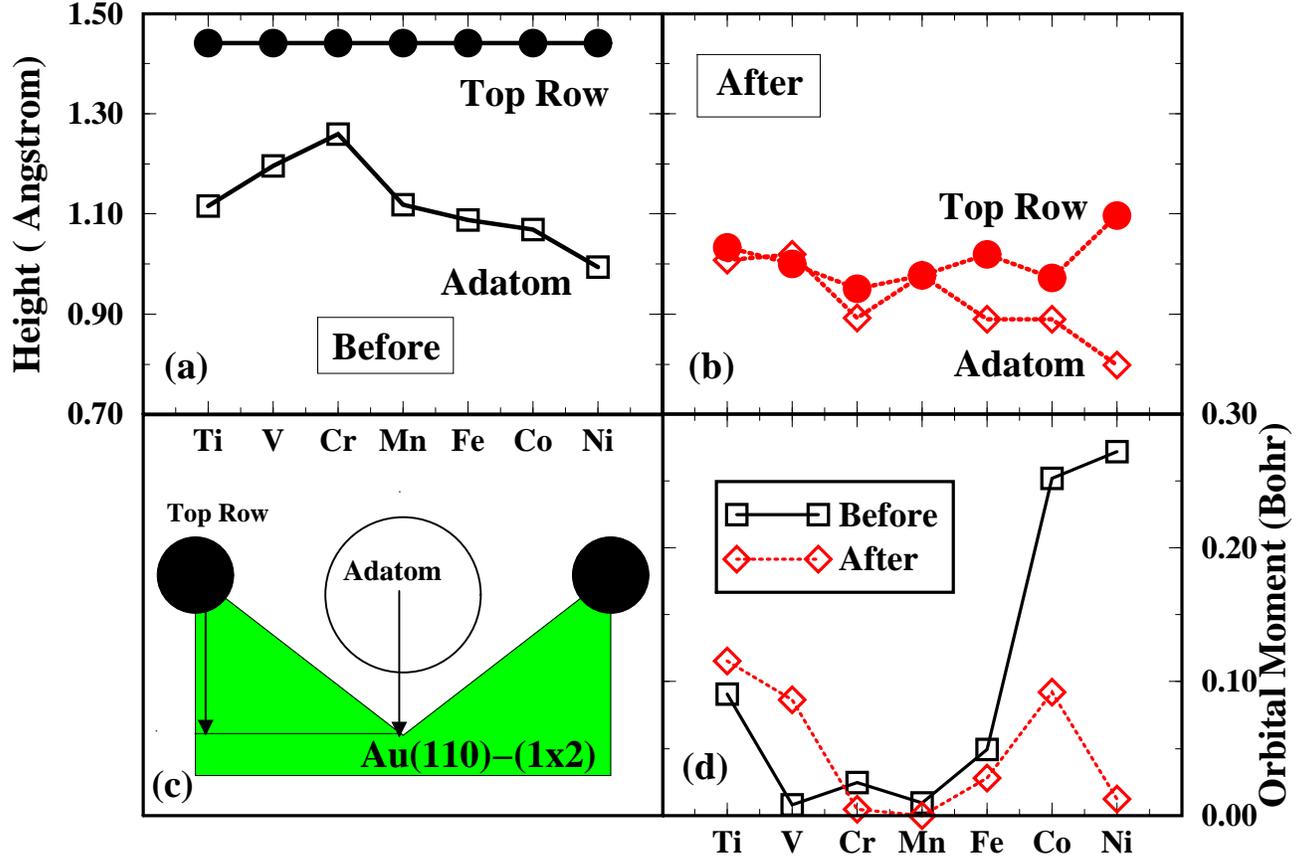}
 \caption{\label{fig3} The effects of structure-relaxation on orbital moments of the absorbed 3d transition metal atoms.
 (a). The heights of top-row and absorbed 3d atoms before the full-structure relaxations.
 The stable positions are different for different 3d atoms, which are obtained by the single-atom
 optimizations on the unrelaxed surface. (b). The heights of top row and absorbed 3d atoms after
 having done full-structure relaxations. The V atom is higher above the top row after the full-relaxation.
 (c). The schematic figure defines the heights of top row and absorbed atoms using two arrows pointing to
 the bottom of the trough. (d). The orbital moments change before and after the full-relaxations.}
 \end{center}
 \end{figure}

 The coordination number of surface atoms is generally smaller than those in bulk. So the magnetism
 of surface is generally stronger than that in bulk. The supported atoms on the surface are in the
 environment similar to the surface atoms. Based on the same logic, the supported atoms on the
 surface have possibly strong magnetism.
 The lack of the orbital polarization isn't the obstacle to study the structural influence on
 the orbital moments. In order to clearly illustrate the effects of the crystal deformations (or the
 cubic distortions), we compare the results on the deformed surface with that on the prefect
 Au(110)-(1$\times$2) surface. The absorbed 3d atoms on the prefect surface still modify their positions
 to reach their stable positions, while the surface atoms are fixed. We find from Fig.~\ref{fig3}
 that the surface relaxations generally quench the orbital moments of the absorbed 3d atoms except
 for the absorbed V and Ti atoms. The changes of the orbital moments in response to the relaxations are
 closely related to the changes of the depth of the 3d atoms embedded in the trough of the reconstructed
 Au(110) surface. The absorbed atoms are deeper in the through, their orbital moments are quenched to
 smaller values due to the stronger crystal field. The absorbed V atom rises above the top row of the
 trough after the relaxations and its orbital moment enhances, opposes to the orbital quenching for
 the other 3d atoms except for the absorbed Ti atom. This is due to the weaker crystal field above the
 surface than that in the trough. The absorbed Ti atom has almost the same height as the top row after
 the relaxations and its orbital moment slightly increases. Thus our results indicate that the surface
 relaxation decreases the orbital moments of the absorbed 3d atoms with the excess half-filled 3d states
 and increases the orbital moments of the absorbed atoms with less half-filled 3d states on
 Au(110)-(1$\times$2) surface.

 The 3d density of states of the absorbed 3d atoms on the two surfaces are shown in Fig.~\ref{fig4}.
 The absorbed Ti, V, Co atoms on the Au(110)-(1$\times$2) surface and V, Co atoms
 on the Au(111) surface have large 3d densities of states at the Fermi energies and large orbital moments too. We get the
 same arguments on the magnetism of Ni$_{n}$ (n=1-5) chains. Large density of state $N(E_{f})$ at Fermi
 energy is advantaged to the formation of the orbital moments. The single Ni atom has small magnetic
 moment, however we can see in section \ref{Ni_Chain} the Ni clusters such as one-dimensional Ni
 chains have significant magnetic moments on the Au surface. The exchange interactions between Ni atoms
 prevent from the decay of magnetic moments of Ni atoms.

 \section{\label{Ni_Chain} The magnetism of Ni atomic chains}

 The initial Ni chains are located in the trough along the A-[1$\bar{1}$0] direction with nearest-nearby
 distance about 2.88$\AA$ the same as the nearest-nearby distance of the missed Au
 chain [Fig.~\ref{fig1}]. The surface slab includes 4 atomic layers with 56 Au atoms and the size of the
 super-cell is 23.046$\AA$$\times$8.148$\AA$$\times$15.52$\AA$. We have relaxed the initial
 structures using the conjugate-gradient Methods. The 2$\times$2$\times$1 K-points mesh
 is used to sampling Brillouin-zone in the corresponding electronic-structure calculations. Based on the relaxed
 structures, we calculated their electronic structures more exactly using RMM-DIIS algorithm with the
 (4$\times$4$\times$1) Monkhorst-Pack grids sampling the first Brillouin zone.
 The other information is the same as in the calculations for single 3d atoms.

 We have also calculated the magnetism of free-standing Ni chains by removing all surface atoms but still
 preserving in the same box cell. There
 are mirror atomic chains because of periodic boundary condition along three directions. The interaction of
 free-standing chains with its mirror atomic chains are small and negligible if the box cell is large
 enough. We have only chosen short chains Ni$_{n}$ where $n$ less than 6 and large box size $23 \AA$ along
 the direction parallel the chain so that the ends of chains are far away from the boundaries of the box
 cell. The energetically favorable configuration of one-dimensional free-standing chains has zigzag shapes
 and not straight line. The Ni$_{n}$ atomic chains in the trough of Au(110)-(1$\times$2) surface are  straight. As we optimize the structures of free-standing chains, all atoms are only allowed modified
 their position along the direction of chains and the optimized free-standing Ni chains still keep a
 straight line shape. We try to compare the results of surface-support straight chains with those of
 free-standing straight  chains. The quantization axis is still along the C-[110] direction in this section.

 \subsection{\label{Ni_chain_15} Magnetism of Ni$_{n}$ (n=1-5)}

 All Ni atoms in the cell have their initial magnetic moments in our calculations.
 We calculate the total energies,
 the spin and orbital magnetic moments of the supported Ni chains when they are magnetized along
 A-[1$\bar{1}$0] direction parallel to the chains or C-[110] direction perpendicular to the chains
 respectively. The magnetization of an atomic chain in this paper means that the magnetic moments for
 every Ni atom align along the same direction. The values of initial moments of Ni atoms are all the same,
 however the values after the self-consistent calculations are probably not the same for all Ni atoms.
 The orientations of magnetic moment may manually change to simulate the different magnetizing directions.
 In a single self-consistent calculation, the directions of magnetic moments change very small for the
 ferromagnetic Ni chains although we have used the unconstrained Density Functional calculations
 (see the table.(\ref{table2})).

 The chains become short after the relaxations and the average nearest-nearby distance
 of Ni$_{5}$ chain is about 2.75 $\AA$ shorter than 2.88 $\AA$ that of initial structure [the insert
 figure of Fig.\ref{fig5}(b) and Fig.\ref{fig1}]. We choose two magnetizing directions: A-[1$\bar{1}$0]
 direction parallel
 to the chain and C-[110] direction perpendicular to the chain. The size-dependence magnetic moments are shown in
 Fig.~\ref{fig5} (a). From this figure we can see that the  spin moments of a single absorbed
 Ni atom are small, 0.147$\mu_{_{B}}$ for the magnetization along C-[110] direction and 0.127$\mu_{_{B}}$ for
 A-[1$\bar{1}$0] direction.
 The spin moments pre atom of the Ni$_{2}$ chain increase to 0.455 $\mu_{_{B}}$ for A-[1$\bar{1}$0] direction and
 0.464 $\mu_{_{B}}$ for C-[110] directions. The spin moments of Ni$_{3}$ have slightly decrease for C-[110]
 direction but
 large decrease for A-[1$\bar{1}$0] direction. The spin moments decrease slightly and try to keep a constant as the chain
 length increases above 4. Our results show that the spin moments of Ni atoms on the reconstructed
 Au(110) surface are smaller than the value 0.675$\mu_{B}$ in crystal Ni having been calculated using
 the same methods in our work.

 The magnetic order is an important aspect of material magnetism. Experimentally the long ferromagnetic
 order is found for Co chains stabilized above a finite temperature on Pt(997)
 surface~\cite{Gambardella1}.
 Above the temperature, the long ferromagnetic order is destroyed and changed into the short ferromagnetic
 fragments. In our calculations, the short Ni chains are ferromagnetic. The Fig.\ref{fig6} shows the
 ferromagnetic order of the surface supported Ni$_{5}$ chains for the magnetization for C-[110] and  A-[1$\bar{1}$0] directions.
 The spin-polarizations of surface Gold atoms near the Ni chain are induced by the magnetic Ni atoms, which
 almost disappear for Gold atoms far from the Ni chain.  We have also calculated the
 anti-ferromagnetic configurations, non-magnetic configurations and other non-collinear magnetic
 configurations and found their energies higher than that of the ferromagnetic configuration. Thus our
 calculations show that the Ni chains with ferromagnetic order are most energetically favorable. It is
 interesting to find from this figure that, when magnetizing the Ni chains parallel to the chain, the  orbital moments are larger than that perpendicular to the chains.
 The spin moments for every Ni atom of the ferromagnetic Ni chain almost don't change their directions
 after the Kohn-Sham self-consistent calculations for all three magnetization directions
 [table.~(\ref{table2})]. The spin moment of the middle atom is smaller than that of other atoms and
 thus the short Ni chain isn't in prefect ferromagnetic order. The larger spin-moments for atoms at two
 ends are because of the smaller coordination number compared with the middle atoms.

 The interaction between supported Ni$_{n}$ chain and surface can be measured with the interaction
 energy $E_{I}=E^{tot}-E^{s}-E^{f}$ where $E^{tot}$ is the total energy, $E^{s}$ the energy of slab with
 relaxed surface and $E^{f}$ the energy of free-standing chain. We generally calculate the interaction
 energy per atom, which removes the effects of simple summation. The interaction energies are generally
 negative.  Fig.~\ref{fig5}(b) shows that the interaction energies of the Ni chains with the surface
 increase with the chain lengths. In our classical molecular dynamics simulation, we have found the
 oscillation of interaction energy for longer chains induced by lattice
 misfit~\cite{Fan1}.
 These results indicate that short Ni chains are more energetically favorable on Au(110)-(1$\times$2)
 surface. This is consistent with the experiments~\cite{Hitzke1} in which most of Ni chains
 are shorter than 6. Additionally, we calculate the density of states of the Ni$_{n}$ (n=1-5) chains.
 From the Fig.~\ref{fig7}, we can see that the densities of states of 3d states of Ni$_{5}$ chain
 changes very small when the magnetization changes from C-[110] direction to A-[1$\bar{1}$0] direction and
 the spin magnetic moment does also change very small [Fig.~\ref{fig7}]. From the figure~\ref{fig5}, the
 changes of orbital moments are more significant if the magnetizing directions change. Although the orbital
 moments are underestimated due to the absence of (OP) orbital polarization term, our results still show  that magnetic anisotropy are closely related to the orbital degree of freedom. For compact magnetic islands
 on metal surface
 such as Co/Pt(111)~\cite{Gambardella4}, the orbital moments decrease with increasing island sizes. This is
 because only atoms at the edge of island have significant contributions to the total orbital moments.
 However, for one-dimensional island all atoms have significant contribution to the total orbital moments.
 The orbital moments per atom keep constant as the lengths of chains beyond 4.

 \subsection{\label{Ni_chain_aniso} The magnetic anisotropy of Ni$_{5}$ chain}

 In the above subsection we have chosen two magnetization directions A-[1$\bar{1}$0] and C-[110].
 In this subsection, we study the magnetic anisotropy of the Ni$_{5}$ chain and show that C-[110]
 direction is along the easy magnetization axis. The magnetic anisotropy energies
 in S$_{1}$ and S$_{2}$ planes are defined as $\delta E_{1}=E^{C}-E^{A}$ and  $\delta E_{2}=E^{C}-E^{B}$
 respectively. $E^{A}$, $E^{B}$ and $E^{C}$ are the total energies when magnetizing the chain along
 A-[1$\bar{1}$0]
 ($\theta=0^{\circ}$), B-[001] ($\phi=0^{\circ}$)and C-[110] ($\theta=90^{\circ}$)
 directions~\cite{Force1}.
 The magnetic anisotropy
 of one-dimensional structure is more prominent than compact structures such as surface and bulk.
 Fig.\ref{fig8} shows the total energies, the spin and orbital moments change with two magnetization angles
 in above two planes. We find from Fig.~\ref{fig8}(a,b) that there are weak magnetic anisotropy in S$_{1}$
 plane and relative strong the magnetic anisotropy in S$_{2}$ plane. The magnetic anisotropy energies are
 0.158(meV/atom) and 0.524(meV/atom) respectively in the two planes. The shape anisotropy determines the
 large difference of magnetic anisotropy energy in two plane because of one-dimension characteristic of
 mono-atomic chain. In the same plane such as the plane S$_{1}$, the magnetic anisotropy is determined
 by spin-orbit coupling. The easy magnetization directions in both S$_{1}$ and S$_{2}$ planes are all
 along C-[110] direction ($\theta = 90^{\circ}$ and $\phi = 0^{\circ}$), which is perpendicular to the
 chain and the surface. The magnetic anisotropy energies can be represented as  $K_{0}+K_{1}sin^2(\theta)$
 in S$_{1}$ plane and $K'_{0}+K'_{1}sin^2(\phi)$ in S$_{2}$ plane~\cite{Autes1,Brooks2}. The parameters
 $K_{0}$=0.158 (meV), $K_{1}$=-0.158 (meV) in S$_{1}$ and $K'_{0}$=0.0 (meV), $K'_{1}$=0.62 (meV) in S$_{2}$
 are obtained by fitting the magnetic anisotropy curves using above two functions.  The magnetic anisotropy
 isn't good fit with $K_{0}+K_{1}sin^2(\theta)$ in S$_{1}$. The underlaying reasons  are (1) there isn't simple
 cubic symmetry in the region where the one-dimensional atomic chain absorbed; (2) the heterogeneously
 distributed the nearest nearby distances along the chain induced by the lattice misfit between the chain
 and the substrate. On the other hand, the value of $K_{1}$ is close to the lowest limit of allowed error in
 total energy about $\pm$0.1meV, thus more accurate calculations are required. Our results also show that
 $K_{1}$ (or $K'_{1}$) isn't equal to the simple $\delta E_{1}=E^{C}-E^{A}$ (or $\delta E_{2}=E^{C}-E^{B}$).

 Experimentally, the values of magnetic anisotropy energy are from 1 meV
 to 10 meV for magnetic clusters on metal surface and from 0.1 mV to 1.0 meV for magnetic film. Based on our
 calculations, the magnetic anisotropy energies for Ni chains on the reconstructed Au(110) surface are close
 to the values of magnetic film. This is because the atoms on Ni chain replace the atoms of first layer of
 original surface and have almost the same coordinate number in their monolayer film. The experiment in
 Co/Pt(997) has shown that the anisotropy energy about 2 meV for the monatomic Co chains and 0.14 meV
 for the monolayer Co film on Pt(997) surface~\cite{Gambardella1}.
 From Fig.~\ref{fig8}(e) we can find that the orbital
 moments reach the minimum when magnetizing the Ni chain along the easy magnetization direction.
 Fig.~\ref{fig8}(f) also shows that the orbital moments almost keep a constant when the magnetization angle
 $\phi$ changes in S$_{2}$ plane. The directions of orbital moments for every Ni atoms on the chain
 are almost parallel to their spin moments when the magnetizations are along A-[1$\bar{1}$0] and C-[110] axes,  or anti-parallel to their spin moments for the magnetization along B-[001] axis.
 From the curve of magnetic anisotropy energy in $S_{2}$ plane in Fig.~\ref{fig8}(b)
 it's most energetically unfavorable when the spin moment of the Ni chain points
 to B-[001] direction ($\phi=90^{\circ}$).

\subsection{\label{Ni_chain_free} The magnetic anisotropy of free-standing Ni$_{5}$ chain and Discussion}

 In order to illustrate the surface influence on the magnetic anisotropy, as a comparison, we also calculate
 the magnetic anisotropy of free-standing Ni$_{5}$ chain in the same box. All calculating details are the
 same as those of the surface-supported Ni$_{5}$ chain except for removing the surface atoms. The average
 nearest-neighbor distance is about 2.196 $\AA$ shorter than 2.77 $\AA$ for the surface-support Ni$_{5}$
 chain and 2.5 $\AA$ in Ni crystal. The changes of densities of states of 3d states for free-standing
 Ni chain are that the majority-spin-up densities of states are far from to the Fermi Energy.
 The larger densities of states at Fermi energy contributed by the minority-spin-down DOS indicate
 that there are larger spin and orbital moment magnetic moments for free-standing Ni chain.

 We only consider the changes of the total energies with the angle
 $\theta$. From the Fig.~\ref{fig8} (g) we can see that the easy magnetization direction is now parallel to
 the chain which is different from perpendicular to the chain for the surface-supported Ni$_{5}$ chain.
 For free-standing chain we find $K_{0}$=0.0 (meV) and $K_{1}$=8.4 (meV) in S$_{1}$. The anisotropy energy
 is large compared with the surface-supported Ni$_{5}$ chain. The change of the sign of $K_{1}$
 compared with surface-supported chain means that the easy magnetization axis change in S$_{1}$ plane.
 The change of the easy magnetization axis had been found in other theoretical works such as Co chains on
 Pd(110) surface~\cite{Dovila1}. Experimentally, the changes of easy magnetization axis are found in the
 growth of magnetic film on metal surface when the thickness of film is beyond a critical value $d_{c}$.
 The transitions from in-pane to perpendicular plane for Ni/Cu(100)~\cite{Schulz5} and inversely from
 perpendicular to in-plane for Fe/Ag(100)~\cite{Qiu5} and Co/Au(111)~\cite{Allenspach5} are found at
 critical coverage about 5-7 ML. The transitions are close related with the transitions from films
 with heterogeneously distributed strains to films with homogenously distributed strains once beyond
 critical thickness. The atomic chains in this work are found in initial stage of film growth and have
 the easy magnetization axis perpendicular to the surface. Whether the easy magnetization axis changes after  a monolayer had completely grown on the surface is interesting topic in our following work.

 The spin moments keep almost constant when the magnetization angle changes from $\theta=0^{\circ}$ to
 $\theta=90^{\circ}$. The spin moment about 1.14$\mu_{B}$ per atom is significantly large than its values
 in Ni crystal 0.675$\mu_{B}$. The behavior of the orbital moment is similar to the surface-supported  Ni$_{5}$ chain, that is, reaches the minimum when perpendicularly magnetizing the chain. The underlaying reason is
 geometric, that is, the orbital currents (or Molecular Currents) of neighbor atoms have large overlapping
 and their cancelations are also large which is disadvantageous to form larger orbital moments. The values
 of orbital moments from 0.2$\mu_{B}$ to 0.6$\mu_{B}$ are larger one power of the values of surface  supported Ni atoms. The easy magnetization axis of free-standing Ni chain A-[1$\bar{1}$0] is different from the easy
 magnetization axis of surface-supported chain C-[110]. The orbital moment reaches its maximum along the  easy magnetization axis A-[1$\bar{1}$0], $\theta=0^{\circ}$ for free standing chain, which is consistent with  the results for magnetic monolayer,~\cite{Solovyev1,Bruno1} but reaches the minimum along the easy magnetization   C-[110] axis  $\theta=90^{\circ}$ for the surface-supported chain. Thus our results show that the surface
 effects change the easy magnetization axis and decrease both the magnetic anisotropy energy and
 the magnetic moment of free standing chains because of the increase of coordination number
 for every atom on a surface.

\section{\label{conclusion} Conclusion }

 Based on the Density Functional Theory, we have calculated the magnetic anisotropy of the supported Ni
 chains and free-standing chain. The easy magnetization axis of the supported Ni chain is perpendicular
 to the chain and surface but parallel to the chain for free standing chain. The different
 easy magnetization axis for the supported and free standing chain is also found in density functional
 calculations of other system. Our results indicate the closed relationship between the magnetic anisotropy
 and orbital moments of the Ni chains. In the environments of crystal, the orbital moment is generally
 quenched to small value due to the splitting of crystal field. Compared with spin moment, orbital moment
 is more sensitive to the change of magnetization direction. The orbital quenching changes with the change
 of magnetized directions. The magnetic anisotropy of materials are determined both by the orbital quenching
 and spin-orbit coupling. For a single absorbed atom, the surface relaxations generally deform the surface
 and modify the positions of the absorbed atoms. If the relaxations lift the absorbed atom out of the trough
 on the constructed Au(110) surface, the orbital moments enhance due to weaker crystal fields and if the
 relaxations make the absorbed atoms deeper embedded in the trough the orbital moments quench to smaller
 values due to the stronger crystal fields.

\begin{center} {\LARGE \bf Acknowledgement} \end{center}
 The one of authors (W. Fan) is greatly indebted to Prof. Q. Q. Zheng, Dr. J. L.Wang,
 Prof. L. J. Zou and Prof. Z. Zeng for their helpful discussions. This work
 run on computers in Center for computational science, Hefei Institutes of Physical
 Science and on LSSC in Institute of Computational Mathematics and Scientific/Engineering
 Computing, Chinese Academy of Sciences. X. G. Gong is additionally supported by the National
 Science Foundation of China, the special funds for major state basic research and CAS projects
 and Knowledge Innovation Program of Chinese Academy of Sciences under KJCX2-SW-W11.

\newpage

\begin{table}[h]
 \caption{\label{table1}The Wigner Seitz Radius-WSR ($\AA$) and Energy-Cutoff of
 plane wave-Ecut (eV) for PAW potentials of 3d transition metal T$_{3d}$=V, Ti, Cr, Mn, Fe, Co, Ni,
 Cu.~\cite{VaspPaw} The real energy cutoff for T$_{3d}$/Au is
 Ecut(T$_{3d}$/Au)= Max [Ecut(T$_{3d}$),Ecut(Au)].}
 \begin{center}
 \newcommand{\m}{\hphantom{$-$}}
 \newcommand{\cc}[1]{\multicolumn{1}{c}{#1}}
 \renewcommand{\tabcolsep}{0.5pc} 
 \renewcommand{\arraystretch}{1.0} 
 \begin{tabular}{lccccccccc}
 \hline\hline
     & Ti   & V     & Cr    & Mn    & Fe    & Co    & Ni    & Cu   & Au\\
 \hline
 WSR &1.323 & 1.323 & 1.323 & 1.323 & 1.302 & 1.302 & 1.286 & 1.312 & 1.503 \\
 Ecut&178.4 & 192.6 & 227.1 & 269.9 & 267.9 & 268.0 & 269.6 & 273.2 & 229.9
 \\\hline
 \end{tabular}
 \end{center}
 \end{table}

 \vskip 3 cm

 \begin{table}[h]
 \begin{center}
 \caption{\label{table2}The spin and orbital magnetic moments (Bohr) for every atoms of Ni$_{5}$ chain when
 the magnetization directions are along A-[1$\bar{1}$0] ,B-[001] and C-[110] respectively. This table
 shows that the directions of spin magnetic moments are almost unchanged in Kohn-Sham self-consistent
 loop although we have used the unconstrained DFT calculations.}
 \newcommand{\m}{\hphantom{$-$}}
 \newcommand{\cc}[1]{\multicolumn{1}{c}{#1}}
 \renewcommand{\tabcolsep}{0.5pc} 
 \renewcommand{\arraystretch}{1.0} 
 \begin{tabular}{cccccccccccccc}
 \hline\hline
     &\vline& A-[1$\bar{1}$0] &  & &\vline& B-[001] &  & &\vline& C-[110] & & \\
 \hline
     &\vline&M$_{A}$&M$_{B}$&M$_{C}$&\vline&M$_{A}$&M$_{B}$&M$_{C}$&\vline&M$_{A}$&M$_{B}$&M$_{C}$ \\
 \hline
  Ni$_{1}$&\vline& 0.434 & 0.0 & -0.006 &\vline& 0.0 & 0.411 & 0.0 &\vline& ~0.017 & 0.0 & 0.430 \\
  Ni$_{2}$&\vline& 0.449 & 0.0 & ~0.006 &\vline& 0.0 & 0.426 & 0.0 &\vline& -0.012 & 0.0 & 0.449 \\
  Ni$_{3}$&\vline& 0.352 & 0.0 & ~0.000 &\vline& 0.0 & 0.323 & 0.0 &\vline& ~0.001 & 0.0 & 0.366 \\
  Ni$_{4}$&\vline& 0.449 & 0.0 & -0.007 &\vline& 0.0 & 0.452 & 0.0 &\vline& ~0.015 & 0.0 & 0.452 \\
  Ni$_{5}$&\vline& 0.437 & 0.0 & ~0.005 &\vline& 0.0 & 0.446 & 0.0 &\vline& -0.015 & 0.0 & 0.432 \\
 \hline
 \hline
     &\vline&L$_{A}$&L$_{B}$&L$_{C}$&\vline&L$_{A}$&L$_{B}$&L$_{C}$&\vline&L$_{A}$&L$_{B}$&L$_{C}$ \\
 \hline
  Ni$_{1}$&\vline& 0.105 & 0.0 & ~0.000 &\vline& 0.0 &-0.085 & 0.0 &\vline& ~0.009 & 0.0 & 0.080 \\
  Ni$_{2}$&\vline& 0.097 & 0.0 & -0.010 &\vline& 0.0 &-0.082 & 0.0 &\vline& -0.012 & 0.0 & 0.093 \\
  Ni$_{3}$&\vline& 0.092 & 0.0 & ~0.000 &\vline& 0.0 &-0.065 & 0.0 &\vline& ~0.001 & 0.0 & 0.060 \\
  Ni$_{4}$&\vline& 0.097 & 0.0 & ~0.010 &\vline& 0.0 &-0.088 & 0.0 &\vline& -0.012 & 0.0 & 0.096 \\
  Ni$_{5}$&\vline& 0.105 & 0.0 & -0.001 &\vline& 0.0 &-0.092 & 0.0 &\vline& -0.009 & 0.0 & 0.079 \\
 \hline
 \hline
 \end{tabular}
 \end{center}
 \end{table}

 \begin{figure}
 \begin{center}
 \includegraphics{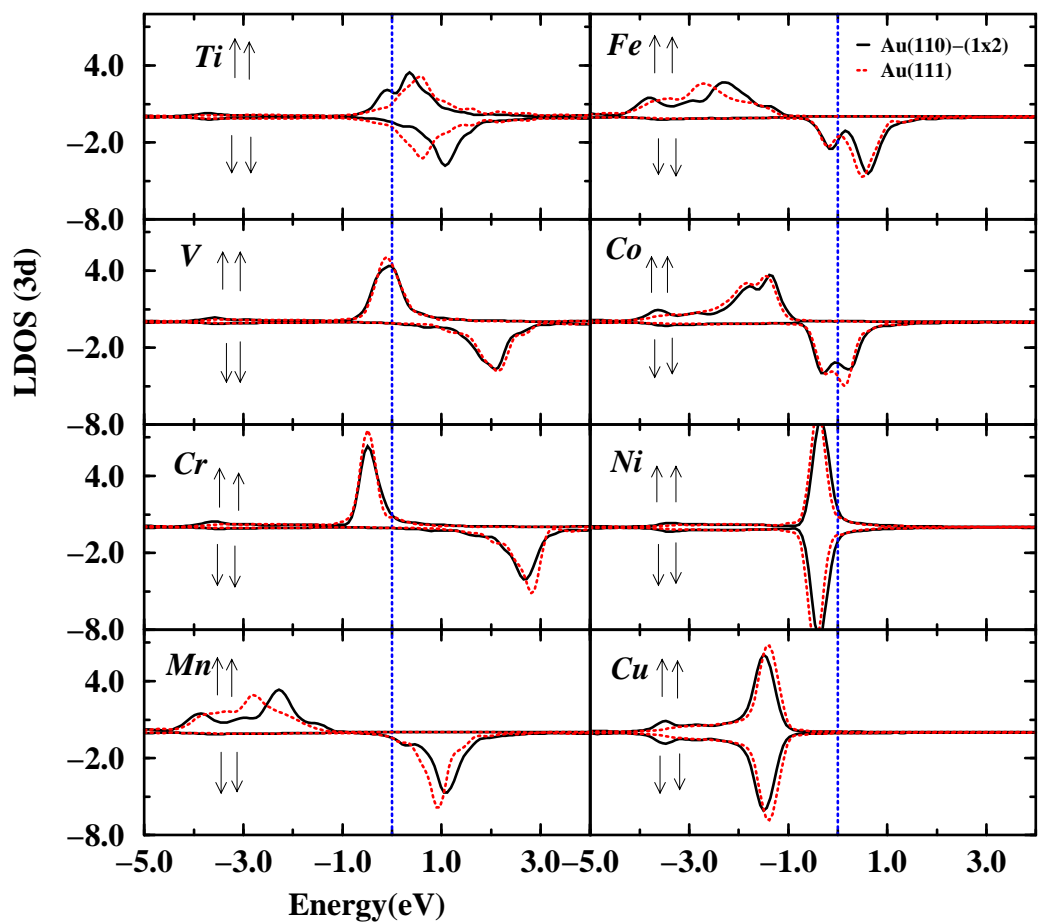}
 \caption{\label{fig4} The 3d densities of state (DOS) for the
 3d atoms absorbed on Au(110)-(1$\times$2) (solid lines) and Au(111) (dot lines )
 surface.}
 \end{center}
 \label{fig4}
 \end{figure}

 \begin{figure}
 \begin{center}
 \includegraphics{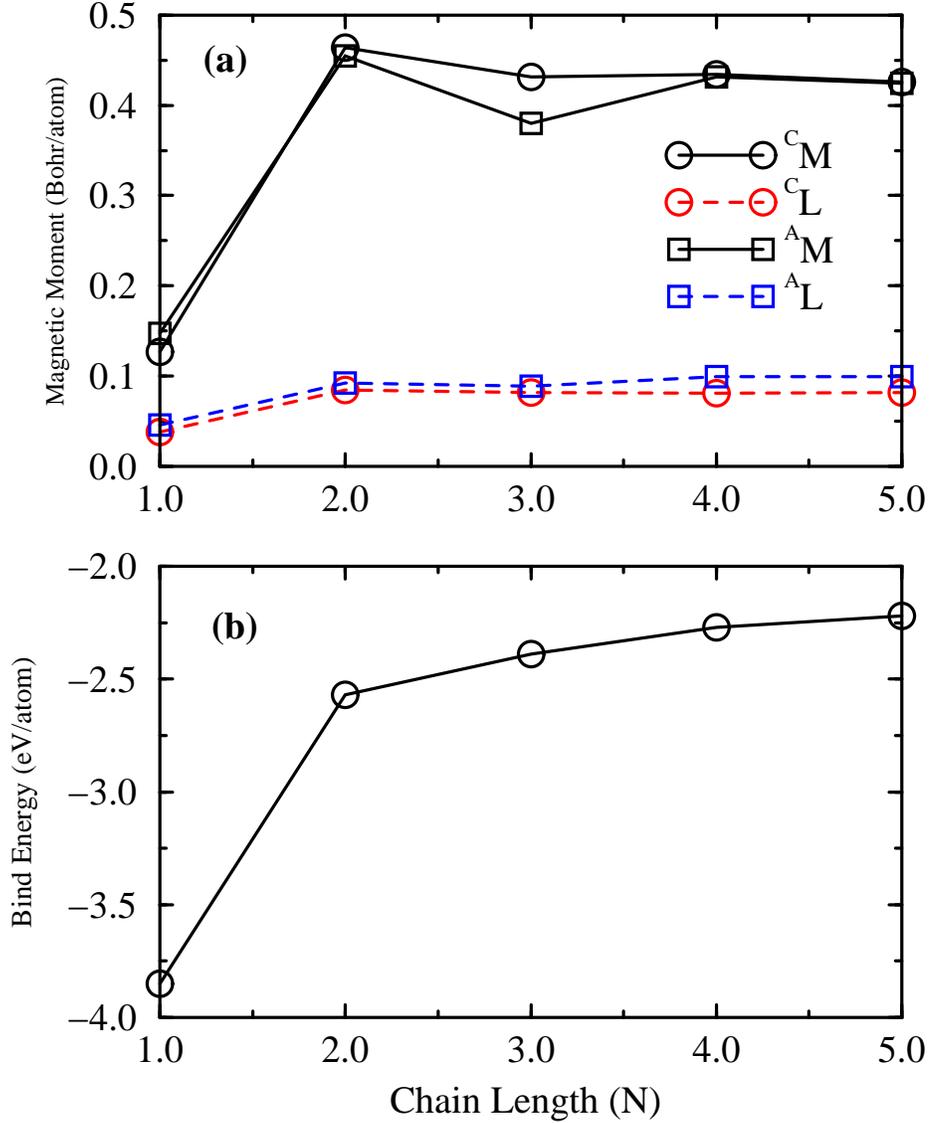}
 \caption{ \label{fig5} (a) The spin $^{A}$M, $^{C}$M and orbital $^{A}$L, $^{C}$L  moments change with the
 length of Ni chains. $^{A}$M and $^{C}$M aren't the A and B  components of the spin moment $\vec{M}$. They
 are the spin magnetic moments when the magnetization along the A-[1$\bar{1}$0] and C-[110] direction and  are
 obtained from different self-consistent  calculations. $^{A}$L and $^{A}$L have the same meaning for
 orbital moments.  (b) Interaction energies between the Ni chains and the surface. The larger absolute values
 indicate the stronger interaction with the surface. }
 \end{center}
 \end{figure}

 \begin{figure}
 \begin{center}
 \includegraphics{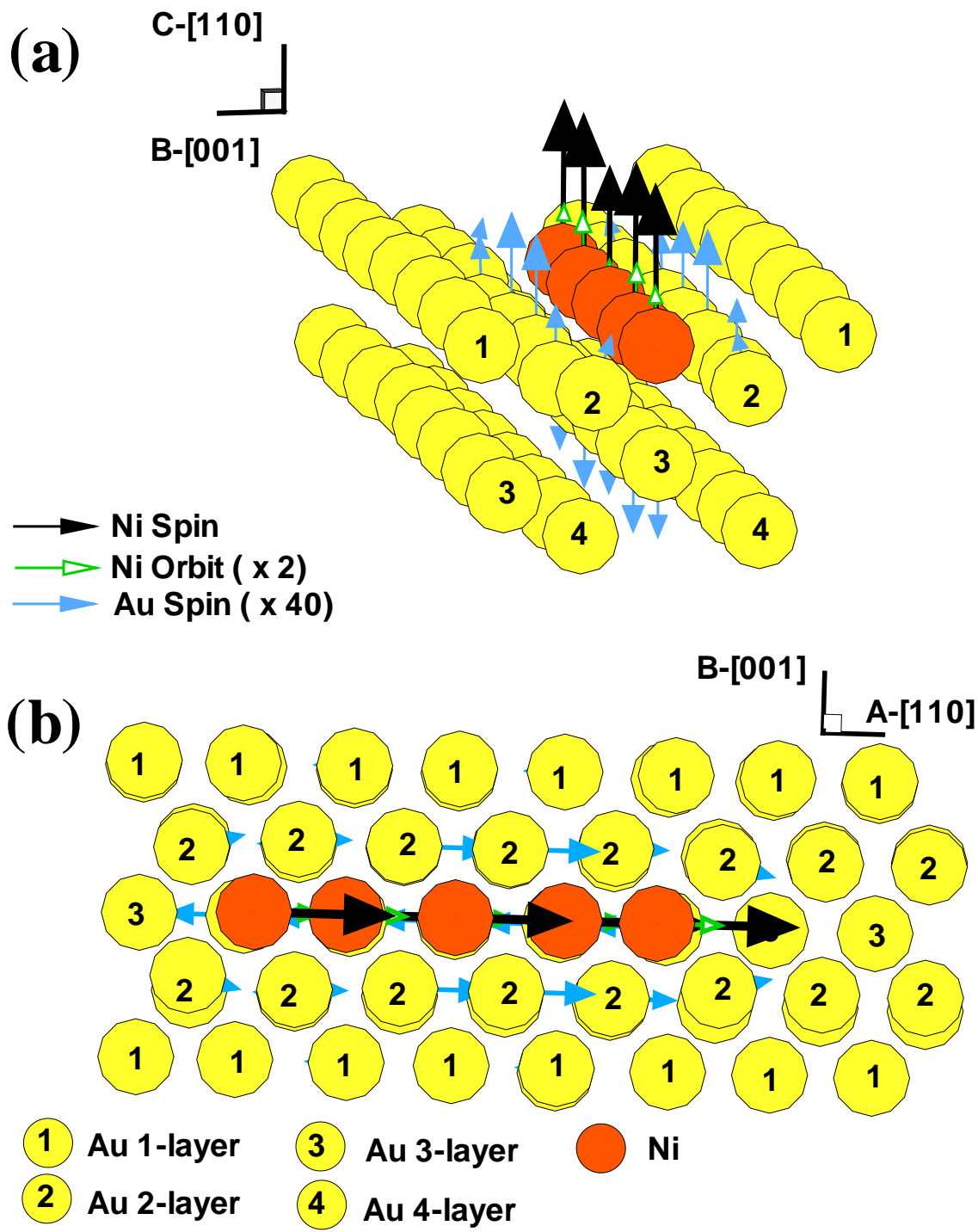}
 \caption{\label{fig6}The vector graph of the spin magnetic moments of the
 Ni chain which illustrates the ferromagnetic order of the Ni$_{5}$ chain based on our calculations.
 Especially, (a) and (b) are corresponding to the magnetization along C-[110] and A-[1$\bar{1}$0]
 direction respectively. The smaller spin magnetic moments of Au atoms and the orbital magnetic
 moments of Ni atoms are multiplied by 40 and 2 respectively to show them with the larger spin
 moments of Ni atoms together in the same figure clearly.}
 \end{center}
 \end{figure}

 \begin{figure}
 \begin{center}
 \includegraphics{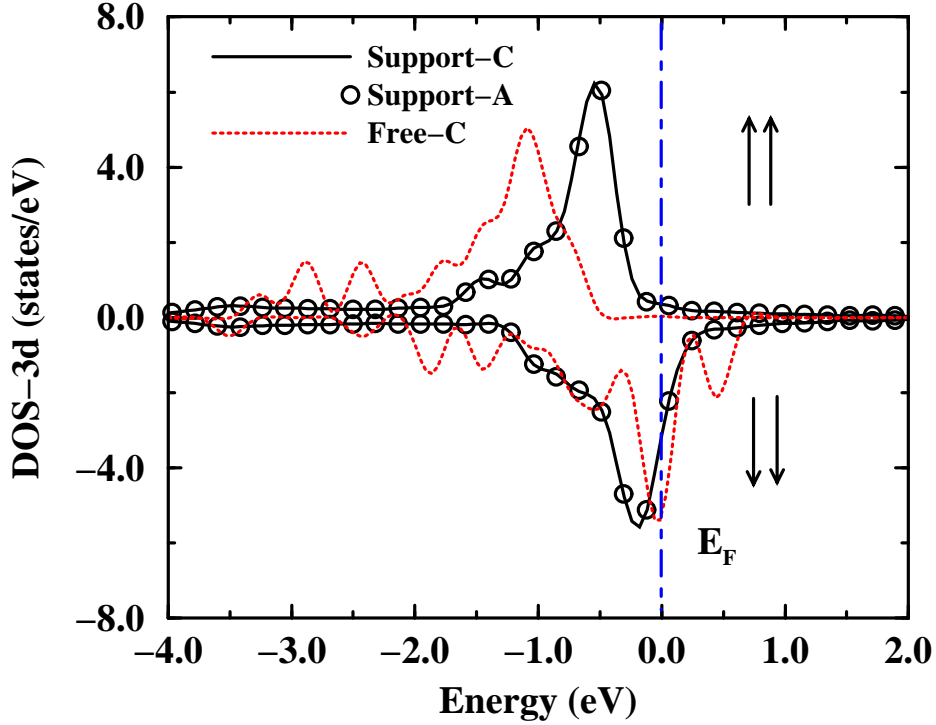}
 \caption{ \label{fig7} The projected DOS of 3d states of the supported Ni$_{5}$
 chain along the directions of magnetization when the chain is magnetized along
 different directions respectively (the solid lines for the C-[110] direction and
 the circles for the A-[1$\bar{1}$0] direction). The spin-up (or spin-down) means
 along the positive (or negative) direction of two magnetization directions. This
 shows that spin degree of freedom is less relevant to magneto-crystalline
 anisotropy. The dots line is the DOS for free-standing Ni chain when magnetizing
 it along C-[110] direction.}
 \end{center}
 \end{figure}

 \begin{figure}
 \begin{center}
 \includegraphics{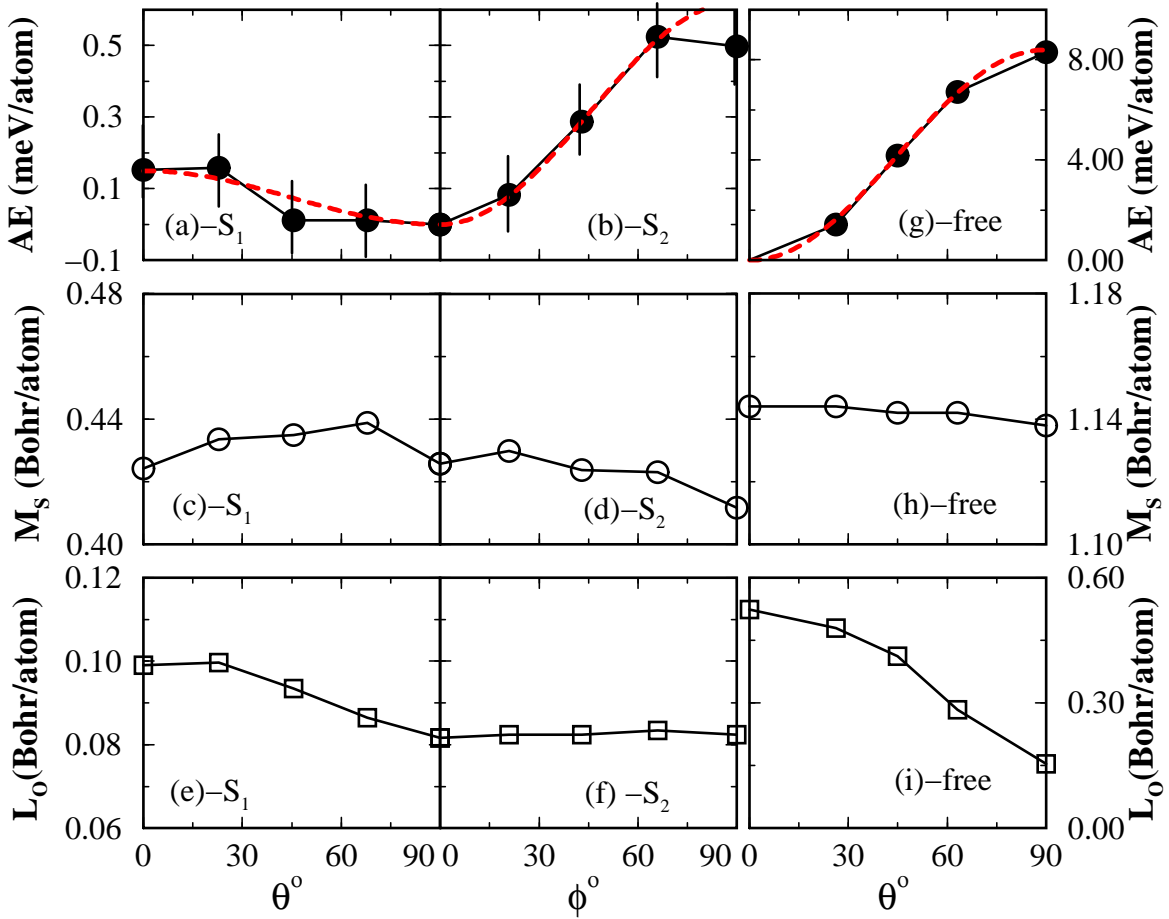}
 \caption{ \label{fig8}
 The anisotropy energies (a,b), the spin (c,d) and orbital moment (e,f) of the supported Ni$_{5}$ chain
 change with the magnetization angles $\theta$ and $\phi$ in the planes S$_{1}$ and S$_{2}$ defined in
 above. The easy magnetization direction is perpendicular to the surface $\theta=90^{\circ}$ and
 $\phi=0^{\circ}$. The anisotropy energies (g), spin moment (h) and orbital moment (i) of the free standing
 Ni$_{5}$ chain change with the magnetization angles $\theta$ in the planes S$_{1}$. The
 easy magnetization direction is parallel to the chain $\theta=0^{\circ}$. Error bars in (a,b) represent
 the lowest limit of allowed error in total energy about $\pm$0.1meV. The magnetic anisotropy energy curves
 are fitting  using function $K_{0}+K_{1}sin^2(\theta)$ for (a,g) and $K_{0}+K_{1}sin^2(\phi)$
 for (b)(red dash-lines). }
 \end{center}
 \end{figure}

\end{document}